\documentclass[preprints,article,accept,moreauthors,pdftex,10pt,a4paper]{mdpi}
\firstpage{1} 
\pubvolume{xx}
\issuenum{1}
\articlenumber{5}
\pubyear{2018}
\copyrightyear{2018}
\history{Received: date; Accepted: date; Published: date}

\pdfoutput=1

\newcommand{\ket}[1]{\vert #1\rangle}

\Title{Dissipative synthesis of mechanical Fock-like states}

\Author{Matteo Brunelli $^{1}$ and Oussama Houhou $^{2,3}$}

\AuthorNames{Matteo Brunelli and Oussama Houhou}

\address{%
$^{1}$ \quad Cavendish Laboratory, University of Cambridge, Cambridge CB3 0HE, United Kingdom\\
$^{2}$ \quad Centre for Theoretical Atomic, Molecular, and Optical Physics, School of Mathematics and Physics, Queen's University, Belfast BT7 1NN, United Kingdom\\
$^{3}$ \quad Laboratory of Physics of Experimental Techniques and Applications, University of Medea, Medea 26000, Algeria}
\corres{Correspondence: mb2236@cam.ac.uk}

\abstract{
The observation of genuine quantum features of nano-mechanical motion is a key goal for both fundamental and applied quantum science. 
To this end, a promising approach is the stabilization of nonclassical features in the presence of dissipation, by means of the 
tunable coupling with a photonic environment. Here we present a scheme that combines 
dissipative squeezing with a mechanical nonlinearity to stabilize arbitrary approximations of (displaced) mechanical Fock state of any number.
We consider an optomechanical system driven by three control lasers---at the cavity resonance and at the two mechanical sidebands---that 
couple  the amplitude of the cavity field to the resonator's position and position squared. 
When the amplitude of the resonant drive is tuned to some specific values,
the mechanical steady state is found in a (displaced) superposition of a finite number of Fock states, which for large enough squeezing 
achieves near-unit fidelity with a (displaced) Fock state of any desired number.}


\begin{document}
\section{Introduction}
The motional state of  atomic or mechanical degrees of freedom can be manipulated via the  interaction with the electromagnetic field confined in a cavity. Such a possibility is best illustrated by cavity cooling, which has been successfully applied to single atoms \cite{AtomCooling}, ions \cite{IonCooling}, and micro- and nano-mechanical resonators~\cite{Chan2011,Teufel2011,Kippen2012}. Recent  breakthroughs in the dissipative preparation of mechanical squeezed states~\cite{MechSqueezing1,MechSqueezing2,MechSqueezing3,ResEngIons3}, where a cavity-assisted scheme is designed to {\em cool} the target system directly into a squeezed state of motion, can be thought of as a powerful development of this paradigm~\cite{ResEng1,Kronwald,Wang,Woolley,JieLi}.  
However, for many applications, ranging from fundamental tests of quantum mechanics to quantum information precessing, the stabilization of highly pure states
with non-Gaussian features is needed instead. 
In cavity optomechanics,  the quadratic optomechanical coupling has been exploited for the dissipative preparation of Schr\"odinger cat states~\cite{OptoCat1,OptoCat2}, but  
the existence of multiple steady states requires the unpractical initialization of the system in a state of definite parity.
Recently we have shown that  a tunable optomechanical coupling which has both a linear and quadratic component enables the stabilization of pure non-Gaussian states without requiring any initialization~\cite{Me1,Me2}. 
For specific values of the amplitude of the laser drives new families of nonclassical states can be stabilized, which correspond to (squeezed and displaced) superpositions of a finite number of Fock states. Here we focus on a specific instance, namely on one such (displaced) finite superposition that  approximates---in principle with arbitrary fidelity---any number state in the harmonic ladder (modulo a displacement). 

\section{Results}

We consider an optomechanical system where the frequency of a cavity mode 
parametrically couples to the displacement and squared displacement of a mechanical resonator.
The Hamiltonian is given by (we set $\hbar=1$ throughout)
\begin{equation}
\hat H=\omega_c \hat a^\dag \hat a+\omega_m \hat b^\dag \hat b - g_0^{(1)}\hat a^\dag \hat a (\hat b+\hat b^\dag)- g_0^{(2)} \hat a^\dag \hat a (\hat b+\hat b^\dag)^2 +\hat H_{\mathrm{drive}}\label{HInt}\,,
\end{equation}
where $\hat a$ ($\hat b$) is the annihilation operator of the cavity (mechanical) mode of frequency $\omega_c$ ($\omega_m$) and $g_0^{(1)}$, $g_0^{(2)}$ respectively quantifies  the linear and quadratic single-photon coupling. Such linear-and-quadratic coupling can be realized in
 membrane-in-the-middle setups~\cite{NonLin,NonLin1, NonLin2}, cold atoms~\cite{NonLin3}, microdisk resonators~\cite{NonLin4} and photonic crystal cavities~\cite{NonLin5,NonLin6}. The cavity has a decay rate $\kappa$ and is driven with three tones
\begin{equation}
\hat H_{\mathrm{drive}}=\hat a^\dag\bigl(\varepsilon_- e^{-i\omega_-t}+\varepsilon_0 e^{-i\omega_0t}+\varepsilon_+ e^{-i\omega_+t} \bigr) +\mathrm{H.c.}\, ,
\end{equation}
applied on the cavity resonance ($\omega_0=\omega_c$), and on the lower and upper  mechanical sideband $(\omega_{\pm}=\omega_c\pm\omega_m)$. 
After standard linearization (we dub $\hat d$ the fluctuation operator of the cavity field), moving into a
frame rotating with the free cavity and mechanical Hamiltonian, and focusing on the good cavity limit ($\kappa\ll \omega_m$) we get 
\begin{equation}
\hat H_{{\rm RWA}}=-\hat d^{\dag}(G_- \hat b +G_+\hat b^{\dag}+G_0\{\hat b,\hat b^{\dag} \}) +\mathrm{H.c.}\, ,
\end{equation}
where we set $G_{\pm}=g_0^{(1)}\alpha_{\pm}$, $G_{0}=g_0^{(2)}\alpha_{0}$, and $\alpha_{\pm,0}$ are the steady values of the cavity amplitude at each frequency component; we will assume these couplings to be real and positive without loss of generality.
After a transient time the cavity field is found in the vacuum while the mechanical resonator in a pure state $\ket{\varphi}$ that satisfies the  condition 
\begin{equation}\label{Eqf}
(G_- \hat b +G_+\hat b^{\dag}+G_0\{\hat b,\hat b^{\dag} \}) \ket{\varphi}=0\, .
\end{equation}
Note that when the nonlinear term in absent, namely $\nobreak{G_0\equiv0}$, we recover dissipative squeezing with a squeezing degree $ r=\tanh^{-1} (G_+/G_-)$~\cite{Kronwald}.
\par
In order  to characterize the steady state $\ket{\varphi}$,
let us first assume that the amplitudes at the two mechanical sidebands are equal, i.e., $G_{\pm}=G$. In this case it is enough to notice that for the following values of the resonant coupling
\begin{equation}\label{Value}
G_0=\frac{G}{\sqrt{2(2n+1)}}\, ,
\end{equation}
the condition expressed in Eq.~\eqref{Eqf} becomes 
\begin{equation}
\hat b^{\dag}\hat b \,\hat D\left(\sqrt{n+\tfrac12}\right) \ket{\varphi}=  n \hat D\left(\sqrt{n+\tfrac12}\right) \ket{\varphi}\, ,
\end{equation}
where $\hat D$ is the displacement operator and $n\in \mathbb{N}$ a non-negative integer (to stress this dependence we set $\ket{\varphi_n}\equiv \ket{\varphi}$ from now on). 
This is in turn equivalent to  
\begin{equation}\label{InfiniteSq}
\ket{\varphi_n}=  \hat D \left(-\sqrt{n+\tfrac12}\right) \ket{n}
\end{equation}
and proves that the steady state is indeed a displaced Fock state. In particular, by tuning the amplitude of the resonant drive in 
Eq.~\eqref{Value} {\it any} state in the Fock state ladder can be stabilized. 
\par
\begin{figure*}[t!]
\centering
\includegraphics[scale=.62]{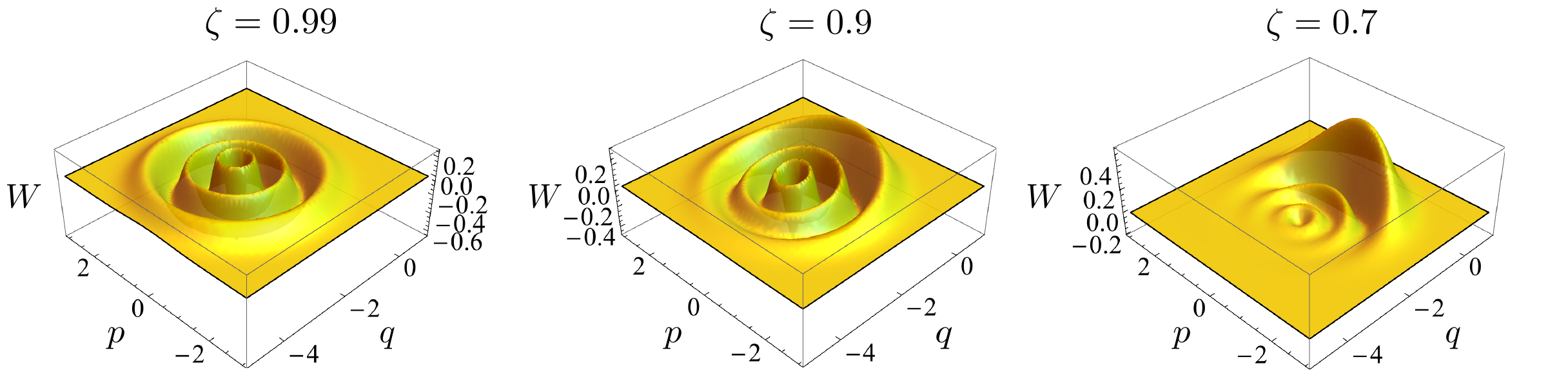}
\caption{Wigner function $\nobreak{W(q,p)=\frac{1}{\pi}\int_{\mathbb{R}}\mathrm{d} y\, e^{2ipy} \varphi_5(q+y)^*\varphi_5(q-y)}$ of the state $\ket{\varphi_5}$ for different values  $\zeta=0.99$ (left), $\zeta=0.9$ (centre), $\zeta=0.7$ (right). 
\label{f:Plot1}}
\end{figure*}

The class of steady states obtained in Eq.~\eqref{InfiniteSq} turns out to be unstable~\cite{Me2}. However, it can be seen as the limit $G_+\rightarrow G_-$ of the more general case $G_+\neq G_-$ with   
\begin{equation}\label{Value2}
G_0=\sqrt{\frac{G_+G_-}{2(2n+1)}}\, \, ,
\end{equation}
which is guaranteed to be stable as long as $G_+<G_-$.
In order to find the new steady state, we can project  Eq.~\eqref{Eqf}
onto the position eigenstate $\ket{q}$ and obtain a differential equation for the associated
wave function $\varphi_n(q)$. The solution of such equation reads
\begin{equation}
\varphi_n(q)\propto e^{-q\sqrt{\zeta(1+2n)}} e^{-\frac{q^2}{2}}H_n\left(q+\tfrac{(1+\zeta)\sqrt{\zeta(1+2n)}}{2\zeta}\right) \, ,
\end{equation}
where we set 
$\zeta=\tanh r\in[0,1)$. 
Note that the integer order of  the Hermite polynomial is determined by the resonant coupling in Eq.~\eqref{Value2}. 
By completing the square in the exponent we get
\begin{equation}
\varphi_n(q)\propto  e^{-\frac12(q-\xi_n)^2}H_n\left(q-\xi_n+\tfrac{(1-\zeta)\sqrt{\zeta(1+2n)}}{2\zeta}\right) \, ,
\end{equation}
where $\xi_n=-\sqrt{\zeta(1+2n)}$. Note that for $\zeta\rightarrow1$ we correctly recover the wave function of a displaced quantum harmonic oscillator. We now exploit the following property of the Hermite polynomials,
$H_n(x+y)=\sum_{k=0}^n \binom{n}{k}H_k(x)(2y)^{n-k}\, ,
$ which leads us to
\begin{equation}
\varphi_n(q)\propto  \sum_{k=0}^n \binom{n}{k}c_n^{-k} e^{-\frac12(q-\xi_n)^2}H_k\left(q-\xi_n\right) \, ,
\end{equation}
with $c_n=-\frac{(1-\zeta)}{4\zeta}\xi_n$. From the last line we can finally read the explicit expression of the state
\begin{equation}
\ket{\varphi_n}=\mathcal{N}_n \hat D\bigl(\xi_n/\sqrt2\bigr) \sum_{k=0}^n \binom{n}{k}c_n^{\,-k} \ket{k}\, ,
\end{equation}
where  the normalization factor is given by $\mathcal{N}_n=\left[_2F_1\left(-n,-n;1;c_n^{-2}\right)\right]^{-1/2}$.
The steady state is now given by the action of a $n$-dependent displacement on a superposition of a finite number ($n+1$) of elements.
It is easily checked that in the limit $\zeta\rightarrow1$ the superposition collapses to the single element of Eq.~\eqref{InfiniteSq}. 
On the other hand, for any non-zero value of the squeezing parameter the state $\ket{\varphi_n}$ displays negativity in the
Wigner distribution and the larger the amount squeezing the closer the resemblance with a Fock state.
This feature is clear from Fig.~\ref{f:Plot1}, where we show the Wigner distribution for a given $n$ ($n=5$) and different values of the squeezing parameter 
$\zeta$. We clearly see that the distribution, which for lower values of $\zeta$ is skewed toward one side, progressively straightens to approach that of a Fock state. We can thus think of $\ket{\varphi_n}$ as a state that approximates any given displace Fock state, to an extent that improves with the amount of  squeezing available.   
Mechanical dissipation---not considered here---sets a limit on the precision of such approximation.
Yet, one can show that it is still possible to approximate with near-unit fidelity any Fock state~\cite{Me2}. 
\par
Coming back to Eq.~\eqref{Eqf}, we notice that $\ket{\varphi_n}$ is the state {\it uniquely} annihilated by the nonlinear operator
\begin{equation}\label{FockMode}
		\hat f=\mathcal{G}\hat \beta+\sqrt{\frac{\cosh r\sinh r}{2(2n+1)}} \{\hat b^{\dag} ,\hat b \} \, ,
	\end{equation} 
	where $\hat \beta=\cosh r \hat b+\sinh r \hat b^{\dag}$ is a Bogoliubov mode and $\mathcal{G}=\sqrt{G_-^2-G_+^2}$. The nonlinear contribution added to the Bogoliubov transformation makes the nature of $\hat f$ non bosonic. 
\section{Discussion}
We presented an exactly solvable model to augment dissipative squeezing by means of a quadratic nonlinearity. The model can be 
implemented in optomechanical cavity and the states stabilized by our protocol approximate displaced multi-phonon Fock state of any desired number.

\acknowledgments{
M.~B.~is supported by the European Union's Horizon 2020 research and innovation programme under grant 
agreement No 732894 (FET Proactive HOT). O.~H.~acknowledges support from the 
SFI-DfE Investigator programme (grant 15/IA/2864), the EU Horizon2020 
Collaborative Project TEQ (grant agreement No 766900) and from the EPSRC 
project EP/P00282X/1.}

\conflictsofinterest{The authors declare no conflict of interest. The funding sponsor had no role in the design of the study; in the collection, analyses, or interpretation of data; in the writing of the manuscript, or in the decision to publish the results.} 

\reftitle{References}





\end{document}